\documentclass{ws-procs9x6-cpt16}
\begin{document}

\newcommand{\refeq}[1]{(\ref{#1})}
\def\etal {{\it et al.}}
\def\al{\alpha}
\def\be{\beta}
\def\ga{\gamma}
\def\de{\delta}
\def\abgd{\alpha\beta\gamma\delta}
\def\ep{\epsilon}
\def\ve{\varepsilon}
\def\ze{\zeta}
\def\et{\eta}
\def\th{\theta}
\def\vt{\vartheta}
\def\io{\iota}
\def\vka{\varkappa}
\def\ka{\kappa}
\def\la{\lambda}
\def\vpi{\varpi}
\def\rh{\rho}
\def\vr{\varrho}
\def\si{\sigma}
\def\vs{\varsigma}
\def\ta{\tau}
\def\up{\upsilon}
\def\ph{\phi}
\def\vp{\varphi}
\def\ch{\chi}
\def\ps{\psi}
\def\om{\omega}
\def\Ga{\Gamma}
\def\De{\Delta}
\def\Th{\Theta}
\def\La{\Lambda}
\def\Si{\Sigma}
\def\Up{\Upsilon}
\def\Ph{\Phi}
\def\Ps{\Psi}
\def\Om{\Omega}
\def\cA{{\cal A}}
\def\cB{{\cal B}}
\def\cC{{\cal C}}
\def\cD{{\cal D}}
\def\cE{{\cal E}}
\def\cl{{\cal L}}
\def\cL{{\cal L}}
\def\cO{{\cal O}}
\def\cP{{\cal P}}
\def\cR{{\cal R}}
\def\cV{{\cal V}}
\def\mn{{\mu\nu}}
\def\prt{\partial}
\def\pt{\phantom}

\def\sb{\overline{s}}
\def\kb{\overline{k}}
\def\mn{{\mu\nu}}
\def\lv{\checkmark}
\def\hs{@{\hspace{22pt}}}
\def\pt#1{\phantom{#1}}
\def\ol#1{\overline{#1}}
\def\prt{\partial}

\title{Looking for Lorentz Violation in Short-Range Gravity}

\author{Rui Xu}

\address{Physics Department, Indiana University,\\
Bloomington, IN 47405, USA}

\begin{abstract}
General violations of Lorentz symmetry can be described by the Standard-Model Extension (SME) framework. The SME predicts modifications to existing physics and can be tested in high-precision experiments. By looking for small deviations from Newton gravity, short-range gravity experiments are expected to be sensitive to possible gravitational Lorentz-violation signals. With two group's short-range gravity data analyzed recently, no nonminimal Lorentz violation signal is found at the micron distance scale, which gives stringent constraints on nonminimal Lorentz-violation coefficients in the SME.
\end{abstract}

\bodymatter

\section{Pure-gravity sector in the SME framework}
Lorentz symmetry is a built-in element of both General Relativity and the
Standard Model. To describe nature using them, we need to 
test this symmetry precisely. Also, if we seek 
a unified theory combining General Relativity and the Standard Model,
we also need to consider possible violations of Lorentz symmetry
that could emerge from the underlying theory,
causing suppressed signals at attainable energy levels.\cite{vs}

The SME framework is an approach to describing Lorentz violation using effective field theory,\cite{dv} where a series of terms that break Lorentz symmetry spontaneously in Lagrange density can be constructed. These terms are couplings between Lorentz-violation coefficients and known fields such as the gravity field, photon field, and fermion fields. For example, in the pure-gravity sector, the Lorentz-violation couplings in the Lagrange density are written as \cite{qvr}
\begin{eqnarray}
\cL_{LV} &=& e \Bigl(\bigl[(k_R)^{\abgd} + (k_R)^{\abgd\la}D_\la + (k_R)^{\abgd\la\si}D_\la D_\si + ...\bigr]R_{\abgd} 
\nonumber\\ 
&&+\bigl[(k_{RR})^{\abgd\mn\ka\rh} + ...\bigr]R_{\abgd}R_{\mn\ka\rh}    +... \Bigr),
\end{eqnarray}   
where $(k_R)^{\abgd}, (k_R)^{\abgd\la}, (k_R)^{\abgd\la\si}, (k_{RR})^{\abgd\mn\ka\rh}, ...$ are Lorentz-violation coefficients. According to the mass dimensions of the coefficients, the term with $(k_R)^{\abgd}$ is called the minimal Lorentz-violation coupling, and all the other terms are nonminimal Lorentz-violation couplings. 

Combining with the Lagrange densities for General Relativity and the Standard Model, as well as taking the fact of spontaneous Lorentz symmetry breaking into consideration, we can formally write down the full Lagrange density for pure-gravity sector SME,\cite{qvr} and hence the modified Einstein field equation.

\section{Weak-field approximation}
The essential problem is that the dynamics of the Lorentz-violation coefficients is unknown. To solve this problem we have to assume both gravity and Lorentz-violation coefficients are weak fields.\cite{qv} Namely, for any Lorentz-violation coefficient $k$, its constant background value $\bar k$ is much larger than its fluctuation $\tilde k$. 

Then, by making the further assumptions that the modified Einstein field equation is diffeomorphism invariant and that the conventional matter energy-momentum tensor is conserved, the leading-order contribution from the dynamics of the Lorentz-violation coefficients is actually fixed with some parameters that depend on the unknown dynamics model.\cite{qvr, qv}

\section{Nonrelativistic solution}
In the weak-field approximation, the modified Einstein field equation turns out to give the modified Poisson equation\cite{qvr, qv}
\begin{equation}
\vec \triangledown ^2 \ph =4 \pi G \rh + (\bar k^{(4)}_{\rm eff})^{jk}\prt_j \prt_k \ph + (\bar k^{(6)}_{\rm eff})^{jklm} \prt_j \prt_k \prt_l \prt_m \ph + ...,
\end{equation}
where $(\bar k^{(4)}_{\rm eff})^{jk}$ is the trace of $(\bar k_R)^{\abgd}$, and $(\bar k^{(6)}_{\rm eff})^{jklm}$ involves more complicated combinations of $(\bar k_R)^{\abgd\la\si}$ and $(\bar k_{RR})^{\abgd\mn\ka\rh}$. Notice the Lorentz-violation coefficients that have odd mass dimensions, for example $(k_R)^{\abgd\la}$, do not appear in the modified Poisson equation. This is a result of conservation of momentum.  

Treating the Lorentz-violation terms in the modified Poisson equation perturbatively, the solution is a modified Newton potential \cite{qvr, qv}
\begin{eqnarray}
\ph  &=& - \frac{GM}{r} \Bigl[ 1 + (\bar k^{(4)}_{\rm eff})^{jk}\frac{r^j r^k}{2r^2}
\nonumber\\ 
&& + \Bigl(\frac{15}{2} (\bar k^{(6)}_{\rm eff})^{jklm} \frac{r^j r^k r^l r^m}{r^6} - 9(\bar k^{(6)}_{\rm eff})^{jkll}\frac{r^j r^k}{r^4} + \frac{3}{2} (\bar k^{(6)}_{\rm eff})^{jkjk}\frac{1}{r^2}\Bigr)  \Bigr].
\end{eqnarray}
which shows violation of rotation symmetry due to the direction dependence, and hence violation of Lorentz symmetry. 
  
\section{Relationship to short-range gravity experiments}
Short-range gravity experiments are designed to test small deviations from Newton gravity at short distance scales, ranging from microns to millimeters in different experiments. By comparing the experiment result with the predicted result from a modified Newton gravity, the parameters in the modified theory can be determined within uncertainties. As the uncertainties are usually larger than the values, the uncertainties are also regarded as constraints on the parameters.
 
In the case of Lorentz-violation gravity, discrete Fourier analysis for the experiment data is required to compare the experimental data and the theoretical result because the Lorentz-violation corrections in the modified Newton potential indicate sidereal-variation signals. The reason for the sidereal variations is that the Lorentz-violation backgrounds $(\bar k^{(4)}_{\rm eff})^{jk}$ and $(\bar k^{(6)}_{\rm eff})^{jklm}$, which are constant in inertial frames
such as the conventional Sun-centered frame,
\cite{vm}
vary in laboratories due to the Earth's rotation.

So far the experimental data analyzed are from the IU and HUST groups.\cite{dvj, jv, cywsjm, cgs} Both groups adopt planar tungsten as test masses. The difference is that the IU experiment detects the force between two test masses, while the HUST experiment detects the torque produced by the force with a torsion-pendulum design. The planar geometry concentrates as much mass as possible at the scale of interest. However, it is insensitive to the $1/r^2$ force. In the modified Newton potential, the $(\bar k^{(4)}_{\rm eff})^{jk}$ term gives a $1/r^2$ force modification. Thus, both IU and HUST experiments are insensitive to $(\bar k^{(4)}_{\rm eff})^{jk}$. As for the nonminimal Lorentz-violation background $(\bar k^{(6)}_{\rm eff})^{jklm}$, combining both experiments gives consistent constraints around $10^{-9}$ m$^2$.

\end{document}